# Improving Motor Imagery Based Brain Computer Interfaces Using A Novel Physical Feedback Technique

Haroun, M. A. and Salah-Eldeen, M.A.*,
* Red Sea STEM - Egypt

*Abstract—* In this project, and through an understanding of neuronal system communication, A novel model serves as an assistive technology for locked-in people suffering from Motor neuronal disease (MND) is proposed. Work was done upon the potential of brain wave activity patterns to be detected as electrical signals, classified and translated into commands following Brain Computer Interfaces (BCI) constructing paradigm. However, the interface constructed was for the first time a device which can reconstruct this command physically. The project novelty is in the feedback step, where an electromagnet's magnetic field is used to showcase the command in ferrofluid droplets movement- these moved to assigned targets due to rotation of a glass surface desk according to the data received from the brain. The goal of this project is to address the challenges of the inaccurate performance in user-training which is yet the main issues preventing BCI from being upgraded into more applicable technology. Tests were performed based on Open ViBE software after uploading recorded files of Motor Imagery MI tasks and the design requirements tested were the motion speed of the droplet and accuracy of hitting fixed targets. An average speed of 0.469 cm/s and average accuracy of 81.6% were obtained from the best volume for the droplet. A conclusion to be drawn was that the promise of this other point of view on BCI systems to be more Brain-Real World Systems.

## I. INTRODUCTION

According to WHO (World Health Organization), about 14% of world's population have disabilities with 2-4% experience significant difficulties in functioning [1]. Motor Neuron Disease (MND) is the name given to a group of neurodegenerative, progressive and incurable conditions that affects the brain and spinal cord. MND is characterized by the degeneration of primarily motor neurons; these neurons fail to work normally and muscles gradually are weakened and waste as neurons degenerate and die. Unable to move, wave or speak, people with MND have only their eyes moving voluntarily [2,3]. To help patients to regain their social life, an alternative way of communication depended on the movement of the eye controls a cursor on a virtual keyboard and selects the desired letters has been around for years [4]. Also, a simple eye blinking was used as a communication method [5]. However, such and similar systems might not be suitable for some patients who have lost the ability to precisely control fine ocular movements or who experience uncontrollable head movements [6,7]. A solution which would allow these patients to communicate is the utilization of modern Brain Computer Interfaces (BCIs).

A BCI system allows people to communicate through brain signals without the need of any muscular movement. In Motor Imagery BCI, users imagine a movement (motor imagery) and the system detects the corresponding spatial distributions of EEG oscillations [8]. However, with an observed potential for BCIs in promoting quality of MND patients' life, there is also a rising need to investigate the possible improvements based on the most suitable use in terms of the user. While the importance of subject's Motor Imagery (MI) skills in BCIs is well recognized, the computer side and improving classification algorithms has been the main area of focus in BCI recent research. [9,10]; powerful machine-learning algorithms have been created. On the other hand, less research has focused on investigating how BCI users may optimally adapt, where the subject performance directly impacts the success of the technology. Moreover, [11,12] have shed light on the route of the latter aspect possible focus. That is, they performed a comparison between feedback modalities. The haptic feedback and visual feedback comparison revealed the rapid improvement of BCI control for the subject when controlled a prosthetic arm compared to the on-screen feedback. This project thus focuses on totally new intervention into the system. More precisely, the goal of this project is to explore the influence of novel intervention of physical feedback design on enhancement of user's performance and interaction with a BCI system.

## II. METHOD

### A. BCI system overview

The BCI system generally performs a five-step process those shown in (figure 1): Acquision, preprocessing, feature extraction, classification, and feedback.

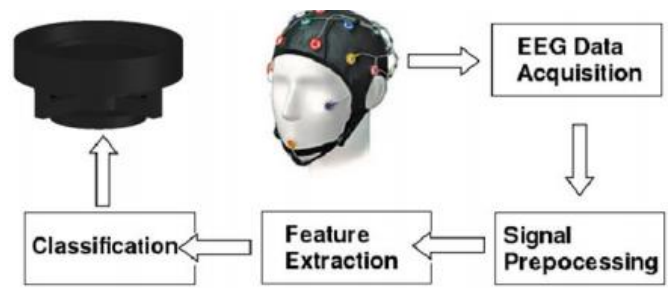

Figure 1: outline of BCI system architecture

EEG, which stands for (electroencephalogram), is a medical record of the electrical activity generated by the brain. The EEG signal is measured as the potential difference over time between signal or active electrode and reference electrode. An extra third electrode, known as the ground electrode, is used to measure the differential voltage between the active and the reference points. The minimal configuration for EEG measurement therefore consists of one active, one reference, and one ground electrode [13].

The purpose of a BCI is to interpret user intentions by means of monitoring cerebral activity. Brain signals involve numerous simultaneous phenomena related to cognitive tasks. Most of them are still incomprehensible and their origins are unknown. However, the physiological phenomena of some brain signals have been decoded in such way that people may learn to modulate them at will, to enable the BCI systems to interpret their intentions. These signals are regarded as possible control signals in BCIs [14,15].. Although there is many kinds of oscillation conditions, like Visually Evoked Potentials (EVP) , Event Related Potentials (ERP) , Readiness Potentials (RP) and P300; what is considered here is Motor Imagery (MI) based oscillations. What makes these neural signal patterns important to human computer interfacing studies is that the mu and beta frequency drop occurs even when the planned movement is not executed—that is, the signals are still observable in paralyzed patients like those with advanced ALS. There by, MI BCI can provide direct communication without any limb movement or external stimulus. MI BCI uses "induced" brain activity from the cortex, rather than "evoked" brain activity [16].

There is a prominent oscillation exhibited in the 8-12 Hz band of the EEG recorded over sensorimotor areas when the subject is not actively engaged in motor action, sensory processing or imagination of actions. This oscillation is usually called mu rhythm and is thought to be produced by thalamocortical circuits. Analysis of mu rhythm showed that it also usually associated with 18-25 Hz beta rhythms. While some of the beta rhythms are harmonics of the mu rhythm, some are separated by topography or timing from mu rhythm. [17].

In the step of acquisition, electrodes of EEG cap are used to obtain the user's imagined limbs movement at specific regions on the scalp. These measurements are amplified electrical signals from the brain neurons and these form the input for a BCI system. The activity is picked by one of different types of electrodes in their most simple method, they are placed on the scalp using gel as a conducting material. The placement of electrodes commonly follows the 10–20 system. The number and location of the electrodes, called EEG channels, are selected according to the paradigm used and mental task performed [18].

After the signals being detected by the headset, There must be filtering of false information such as ECG, EMG, REM, blink and power frequency interference that will interfere with the observation and analysis against the EEG signals by people, this is performed in the preprocessing step [19].

After the preprocessing comes the feature extraction step. It is the process of distinguishing the pertinent signal characteristics from extraneous content and representing them in a compact and/or meaningful form, amenable to interpretation by a human or computer. Classification , after that, is a step which assigns a class label to a set of features extracted from the signal. In other words, it is about assigning a command to a specific pattern of the EEG. The two process are usually integrated in one "processing program" in normal BCI systems. Finally, Feedback is the step of exhibiting the command. It is the step which understood to be revealing the intention of the brain. [20].

*B. OpenViBE Software*

OpenViBE [21] is a software platform serves as an open source, stand-alone software that can be used for quick and robust prototyping of BCI systems. (Figure 2) shows the designer that will be used to develop scenarios that will process, train and classify the EEG signals live from the headset or from previously recorded EEG files.

The experimental motor imagery EEG data, shown in (figure 2) was obtained from the 2005 competition, BCI Competition III Data Iva, published by the German Research Center for BCI movement . The data comes from two healthy subject. During the experiment (in the file uploaded) the subject completed motor imagery of left hand or right hand according to the computer prompts. One subject was asked to perform left hand imagery while the other performed imagery of the right hands. EEG electrodes were placed in the standard 10–20 lead mapping and data collection featured 59 channels with a sampling frequency of 100 Hz.

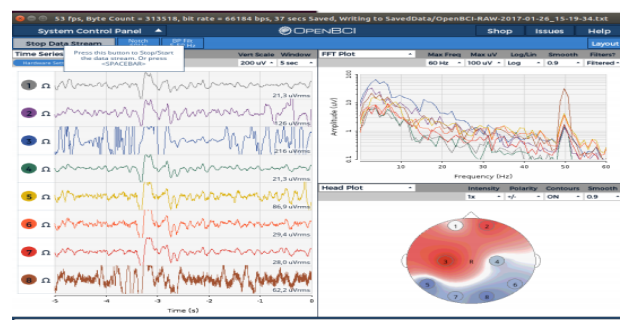

Figure 2: part of working on uploaded file of MI EEG experiment

In the experiment subjects performed 200 motor imagery tasks, each lasting for 8 s. During the task, a cross was first presented for 2 s, indicating that the subject should prepare for motor imagery. Subjects were then shown an arrow pointing left, right, or down for 4 seconds, indicating that the subject should imagine respective motion of the left hand, right hand, or foot. Finally, subjects rested for 2 seconds while a blank screen was shown. The full experimental process can be seen in (figure 3).

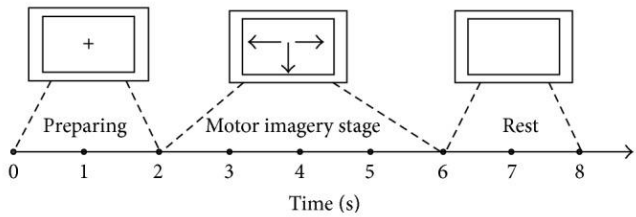

Figure 3: Experiment set of MI EEG subjected to use

*C. Prototype Construction*

Ferrofluid Formation: (figures 4,5) :

There are two basic steps in creating a ferrofluid: synthesis of the magnetic solid or magnetite (Fe3O4), and suspension in water with the aid of a surfactant. The magnetic particles must be very small on the order of 10 nm (100 Å) in diameter so that the thermal energy of the particles is large enough to overcome the magnetic interactions between particles. The magnetite will be synthesized by a precipitation reaction that occurs upon mixing FeCl2 and FeCl3 with ammonium hydroxide (an aqueous solution of ammonia, NH3). The unbalanced equation for this reaction is as follows:

__ FeCl3+ __ FeCl2 + __ NH3+ __ H2O  __ Fe3O4 + __ NH4Cl

The surfactant used in this synthesis is citric acid (C6H8O7). The hydroxide (OH–) ions formed in solution tend to bind to the iron sites on the magnetite particles, creating a net negative charge on each particle. The positively-charged citric ions will then associate with the negatively-charged magnetite particles, forming a kind of shell around each magnetite particle. This charged shell raises the energy required for the particles to agglomerate, stabilizing the suspension.

The experimental steps we followed were:
- Adding 1.0 mL of 2M FeCl2 solution and 4.0 mL of 1M FeCl3 to a 100 mL beaker.
- Then Adding a magnetic stirring bar and begin stirring.
- Using a 50 mL buret to add dropwise 50 mL of 0.7 M aqueous NH3 solution. Magnetite, a black precipitate, will form. Continue stirring throughout the slow addition of the ammonia solution over a period of 5 minutes.
- Turning off the stirrer and immediately use a strong magnet to work the stir bar up the walls of the flask. And then remove the stir bar with gloved hands before it touches the magnet.
- Letting the magnetite settle for a few minutes, then decant (pour off) and discard the clear liquid without losing a substantial amount of solid.
- Transferring the solid to a weighing boat with the aid of a few squirts from a wash bottle.
- Using a strong magnet to attract the ferrofluid to the bottom of the weighing boat.
- Pouring off and discard as much clear liquid as possible. Rinse again with water from a wash bottle and discard the rinse as before.
- Adding 2.0 mL of 25% Citric Acid. Stir with a glass rod to suspend the solid in the liquid. Use a strong magnet to attract the ferrofluid to the bottom of the weighing boat.
- Pouring off and discard the liquid. And finally moving the strong magnet around and pour off any liquid.

The balanced equation of magnetite formation is as follows:

**2FeCl3+ FeCl2 + 8NH3 + 4H2O → Fe3O4 + 8NH4Cl**

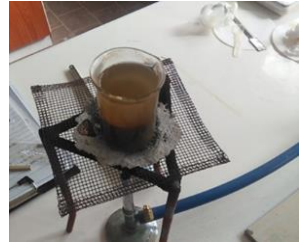 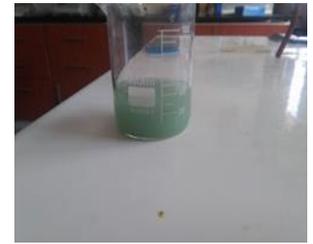

Figure 4: perception of magnetite

Figure 5: solution of ferric chloride

Constructing the controlling system (figure 8,9):

As shown in (figure 6), the interface was designed at AutoCAD and, as shown in (figure 7), then printed by Ultimaker 2+ machine in STEM FabLab. The glass sheet was also shaped to a circular sphere. The radius of the internal circle was 6 cm. There was also another side-connected circle with radius of 7 cm.

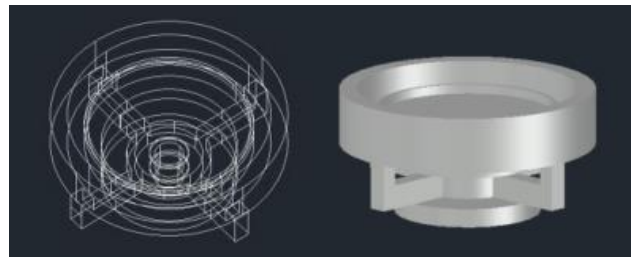

Figure 6: AutoCAD design for the physical interface

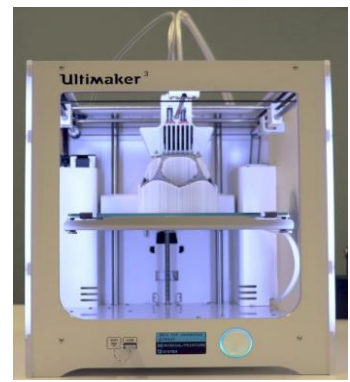

Figure 7: Ultimaker 2+ machine used to construct the designed prototype

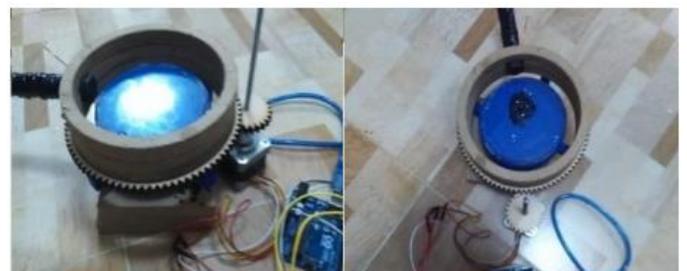

Figures 8,9: Final prototype of the novel physical interface

### D. How The System Works

In the following, mentioned is the novel system to be integrated with the normal BCI systems. To achieve the reliable physical feedback, 3 major concepts was integrated in an intellectually unified paradigm. Those are:

1-Classification software:
The attenuation in mu rhythm which is our feature to be extracted differs between when the imagined motion is to right hand and to left hand. The two different attenuations were given two classes. Classification process begins when this attenuation is detected and then sampled by a MATLAB algorithm. The 2 bit sampling was used to produce the current and then the movement of the ferrofluid based on this difference in the attenuation

2-Arduino board and electromagnet:
The two different 0/1 binaries gathered from the EEG was transformed to an Arduino controlling system. The Arduino was used to initiate a motor to rotate at two different rotation angles. The rotation of the attached desk to the motor has been useful due to the attached electromagnet. That electromagnet was connected with an external source of electricity. Based upon Ampère's circuital law, the electric field in the electromagnet was able to establish a magnetic field. The electromagnet dimensions was estimated as follows: it will be like a solenoid having the length of an iron nail (10cm) with radius of 1cm. According to elementary physics equation which describe the intensity of the magnetic field around the solenoid,

$$B = \frac{\mu n I}{l} = \frac{1.2 \times 10^{-3} \ast 50 \ast 0.01}{5 \ast 10^{-3}} = 0.12\ T$$

3-Ferro fluid droplet movement (figure 10):
The movement of the ferrofluid movement is based, as mentioned before, on the induced current in electromagnet. The idea is that this electromagnet is movable in a circular path based upon a motor and gear. The angle of rotation depends on the integration of the data at the Arduino. Algorithm was used to achieve the following table paradigm.

|  | Binary Code | Angle Of Rotation |
|---|---|---|
| Right Hand MI | 01 | 180 Degrees |
| Left Hand MI | 10 | 360 Degrees |

TABLE 1

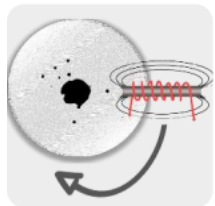

Figure 10: demonstrative design

.

## III. RESULTS AND DISCUSSION

Velocity Test

Velocity is measured by calculating the distance and dividing it by time taken by the droplet to move. Although a force affects the droplet and acceleration should be calculated, High stroke force affect the motion decreasing the magnitude of the acceleration making it negligible. Data is table 2 and figure 1 show this relation such that velocity of the droplet is nearly negligible.

| Distance cm | 20 μL | 10 μL | 5 μL |
|---|---|---|---|
| 3 | 90.98 | 41.98 | 27.13 |
| 6 | 181.95 | 83.79 | 54.65 |
| 9 | 278.81 | 126.93 | 83.47 |

TABLE 2

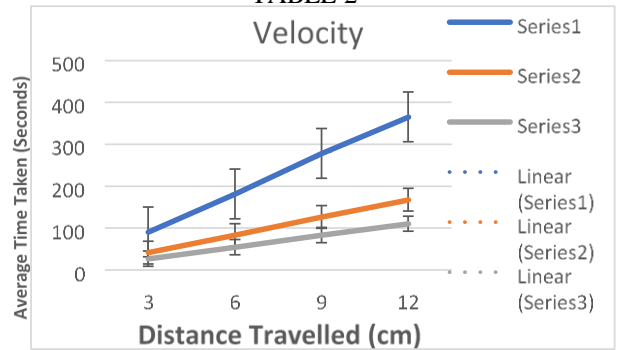

GRAPH 2

| Expected Distance cm | Trial 1 | Trial 2 | Trial 3 | Average | Volume in μL |
|---|---|---|---|---|---|
| 3 | 90.91 | 90.82 | 91.2 | 90.98 | 20 |
| 6 | 181.83 | 182.1 | 181.92 |  | 20 181.95 |
| 9 | 272.91 | 275.71 | 279.12 |  | 20 278.81 |
| 12 | 363.64 | 368.1 | 365.2 |  | 20 365.65 |
| 3 | 41.72 | 42.35 | 41.87 | 41.98 | 10 |
| 6 | 83.54 | 83.62 | 84.21 | 83.79 | 10 |
| 9 | 125.47 | 126.75 | 128.57 |  | 10 126.93 |
| 12 | 166.97 | 168.54 | 167.94 |  | 10 167.82 |
| 3 | 27.54 | 27.39 | 26.47 | 27.13 | 5 |
| 6 | 54.78 | 56.31 | 52.87 | 57.65 | 5 |
| 9 | 81.97 | 84.79 | 83.65 | 83.47 | 5 |
| 12 | 110.91 | 109.87 | 111.54 | 110.77 | 5 |

TABLE 3

The velocity of the droplet ranged from 0.033 cm/s to 0.11 cm/s. Droplets with high volume takes more time to reach a certain distance. Using small efficient volume of ferrofluid, the system can achieve high speed reliability.

Accuracy Test
Accuracy is defined as the how much success does the droplet achieved. A simple ruler was used to determine the actual distance the droplet moved and then compared with the distance the droplet should move. The experiment is considered precise as even the friction is calculated in the stroke force as it defined by the viscosity of the droplet. The distance measurement error by ± 0.2 (20.5%). The measurement tool of distances was the roller. Table 4 shows the average of the many trial tests on accuracy.

| Distance cm | 5 μL | 10 μL | 20 μL |
|---|---|---|---|
| 3 | 80.01% | 82.10% | 69.03% |
| 6 | 81.67% | 78.20% | 59.10% |
| 9 | 81.48% | 70.01% | 62.11% |
| 12 | 83.61% | 76.22% | 55.00% |
| Average | 0.816898 | 0.765278 | 0.613889 |

TABLE 4

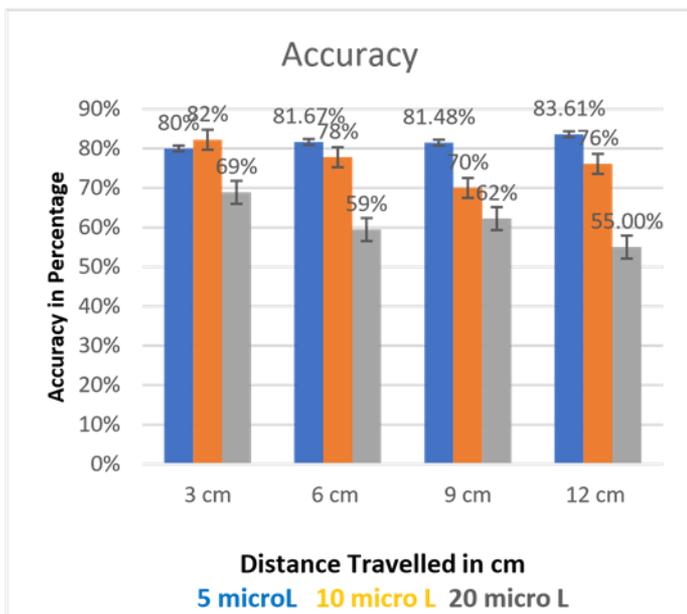

GRAPH 2

Accuracy of achieving the target test results indicated that the volume again should be as small as possible to could be controlled in well manner.

| Time sec. | Trial 1 | Trial 2 | Trail 3 | Average | Volume in μL |
|---|---|---|---|---|---|
| 90.98 | 2.4/ 3 | 2.5 | 2.3 | 2.43 | 5 |
| 181.95 | 5.1/ 6 | 4.7 | 4.9 | 4.91 | 5 |
| 278.81 | 7.3/ 9 | 7.6 | 7.1 | 7.33 | 5 |
| 365.65 | 9.8/ 12 | 10.1 | 10.2 | 10.03 | 5 |
| 41.98 | 2.2/ 3 | 2.3 | 2.9 | 2.467 | 10 |
| 83.79 | 4.4/ 6 | 4.5 | 5.1 | 4.67 | 10 |
| 126.93 | 6.3/ 9 | 6.7 | 5.9 | 6.3 | 10 |
| 167.82 | 8.9/ 12 | 9.1 | 9.4 | 9.13 | 10 |
| 27.13 | 1.7/ 3 | 2.1 | 2.4 | 2.067 | 20 |
| 57.65 | 3.3/ 6 | 3.5 | 3.9 | 3.567 | 20 |
| 83.47 | 5.5/ 9 | 5.1 | 6.2 | 5.60 | 20 |
| 110.77 | 6.4/ 12 | 6.1 | 7.3 | 6.62 | 20 |

TABLE 5

IV. CONCLUSION

There is a major issue that challenge BCI researchers in regard to the human users. It is user training; the operation of a BCI is not intuitive and users need to learn how to voluntarily control their neural activities [22,23,24]. Considering case of motor imagery based BCIs, a rather long training period is required until the users gain skill in the imagery task and achieve optimal performance [25]. Previous studies have sought various kinds of cognitive tasks and feedback techniques to propose an optimal training protocol [24], but no one yet explored changes to the on-screen way of visualizing the command (the feedback). In this research, proved is the possibility of achieving a reliable totally different technique of feedback, and this one is the physical feedback.

As it is the case for new areas research, further research in the field is required to extend the use of our physical visualization system as a wider channel of communication. This can be done by shedding light upon new brain activity patterns in relation with new situations and actions. Extending the system use can be achieved also after increasing the number of options that the interface can manage to exhibit. This can be done by adding more option for the classification and their corresponding command of different angle of rotation. Idea demonstration is shown in figure 10 .(where different colors correspond to different commands)

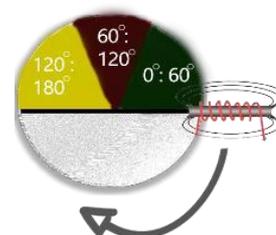

Figure 10 : design for increasing command channels of physical BCI

The tool can also be used for improvement of BCI – games as the tool is suitable to be used ,if other electromagnets were added and a study on the impact of each of which on the other, in games as it would be possible to reconstruct abstract 2D object as shown in figure 11.

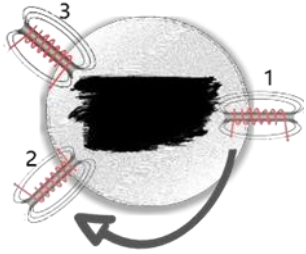

Figure 11 : design for 2D feedback of physical BCI

This multi effect of different magnetic fields can be understood through Maxwell's equations [25]. In our case, we are changing magnetic fields slowly (compared to radio frequencies) thus the magneto-static equations are appropriate. These are

$$\nabla \times \vec{H} = \vec{j}$$
$$\nabla \cdot \vec{B} = 0$$
$$\vec{B} = \mu_o(\vec{H} + \vec{M}) = \mu_o(\vec{H} + \chi\vec{H}),$$

where $\vec{B}$ is the magnetic field [in Tesla], $\vec{H}$ is the magnetic intensity [Amperes/meter], $\vec{j}$ is the current density [A/m2], $\vec{M}$ is the material magnetization [A/m], $\chi$ is the magnetic susceptibility, and $\mu_o = 4\pi \times 10^{-7}$ N/A2 is the permeability of a vacuum. The force on a single superparamagnetic particle is then

$$\vec{F}_M = \frac{2\pi a^3}{3} \cdot \frac{\mu_0 \chi}{1+\chi/3} \nabla \left\| \vec{H} \right\|^2 = \frac{4\pi a^3}{3} \cdot \frac{\mu_0 \chi}{1+\chi/3} \left( \frac{\partial \vec{H}}{\partial \vec{x}} \right)^T \vec{H}.$$

When a magnetic force is applied, a single particle will accelerate in the direction of that force until it sees an equal and opposite fluid (Stokes) drag force. Since the Stokes force is

$$\vec{F}_S = -6\pi a \eta \, \vec{v},$$